\documentclass[aps,floatfix,nofootinbib,showpacs,onecolumn,superscriptaddress]{revtex4}

\usepackage{amsmath}
\usepackage{amssymb}
\usepackage{amsthm}
\usepackage{dcolumn}
\usepackage{epsfig}
\usepackage{graphics}
\usepackage{graphicx}
\usepackage{slashed,epsfig}
\usepackage{longtable}
\usepackage{color}

\definecolor{darkgreen}{rgb}{0,0.5,0}
\definecolor{purple}{rgb}{0.5,0,0.5}
\definecolor{nblue}{rgb}{0.0,0.0,0.50}
\definecolor{scarlet}{rgb}{1.0,0.2,0}
\usepackage[colorlinks=true, pdfstartview=FitV, linkcolor=purple, citecolor= purple, urlcolor=blue]{hyperref}

\newcommand{\beq} {\begin{equation}}
\newcommand{\eeq} {\end{equation}}
\newcommand{\beqa} {\begin{eqnarray}}
\newcommand{\eeqa} {\end{eqnarray}}

\def\beq{\begin{equation}}
\def\be{\begin{equation}}
\def\eeq{\end{equation}}
\def\ee{\end{equation}}
\def\bea{\begin{eqnarray}}
\def\eea{\end{eqnarray}}

\begin{document}

{\par\raggedleft \texttt{SLAC-PUB-14479}\par}
{\par\raggedleft \texttt{CP3-Origins-2011-20 \& DIAS-2011-06}\par}
\bigskip{}

\title{Setting the Renormalization Scale in QCD:\\The Principle of Maximum Conformality}

\author{Stanley~J.~Brodsky}
\affiliation{SLAC National Accelerator Laboratory\\
Stanford University, Stanford, California 94309, USA}
\affiliation{CP$^3$-Origins, University of Southern Denmark\\ Campusvej 55, DK-5230 Odense M, Denmark}
\author{Leonardo Di Giustino}
\affiliation{SLAC National Accelerator Laboratory\\
Stanford University, Stanford, California 94309, USA}
\begin{abstract}
A key problem in making precise perturbative QCD predictions is
the uncertainty in determining the renormalization scale $\mu$ of
the running coupling $\alpha_s(\mu^2).$ The purpose of the running
coupling in any gauge theory is to sum all terms involving the
$\beta$ function; in fact, when the renormalization scale is set
properly, all non-conformal $\beta \ne 0$ terms  in a perturbative
expansion arising from renormalization are summed into the running
coupling. The remaining terms in the perturbative series are then
identical to that of a conformal theory; i.e., the corresponding
theory with $\beta=0$. The resulting scale-fixed predictions using
the  ``principle of maximum conformality" (PMC) are independent of
the choice of renormalization scheme --  a key requirement of
renormalization group invariance.   The results avoid renormalon
resummation and agree with QED scale-setting in the Abelian limit.
The PMC is also the theoretical principle underlying the BLM
procedure, commensurate scale relations between observables, and
the scale-setting method used in lattice gauge theory.  The number
of active flavors $n_f$ in the QCD $\beta$ function is also
correctly determined. We discuss several methods for determining
the PMC scale for QCD processes. We show that a single global PMC scale,
valid at leading order, can be derived from basic properties of
the perturbative QCD cross section. The elimination of the
renormalization scale ambiguity and the scheme dependence using
the PMC will not only increase the precision of QCD tests, but it
will also increase the sensitivity of collider experiments to new
physics beyond the Standard Model.

\end{abstract}

\pacs{11.15.Bt, 12.20.Ds}

\maketitle

\date{\today}

\section{Introduction}

A key difficulty in making precise perturbative QCD predictions is
the uncertainty in determining the renormalization scale $\mu$ of the
running coupling $\alpha_s(\mu^2)$. It is common practice to simply guess
a physical scale $\mu = Q$ of order of a typical momentum
transfer $Q$ in the process, and then vary the scale over a range
$Q/2$ and $2 Q$.  This procedure is clearly problematic since the
resulting fixed-order pQCD prediction will depend on the choice of
renormalization scheme;  it can even predict negative QCD cross
sections at next-to-leading-order~\cite{Maitre:2009xp}.

The purpose of the running coupling in any gauge theory is to sum all
terms involving the $\beta$ function; in fact, when the
renormalization scale $\mu$ is set properly, all non-conformal
$\beta \ne 0$ terms  in a perturbative expansion arising from
renormalization are summed into the running coupling.  The
remaining terms in the perturbative series are then identical to
that of a conformal theory; i.e., the theory with $\beta=0$.  The
divergent ``renormalon" series of order $\alpha_s^n \beta^n n! $
does not appear in the conformal series. Thus  as in quantum electrodynamics, the
renormalization scale $\mu$ is determined unambiguously by the
``Principle of Maximal Conformality (PMC)".  This is also the principle
underlying BLM scale setting~\cite{Brodsky:1982gc}

It should be recalled that there is no ambiguity in setting the
renormalization scale in QED. In the standard Gell-Mann--Low
scheme for QED, the renormalization scale is simply the virtuality
of the virtual photon~\cite{GellMann:1954fq}. For example, in
electron-muon elastic scattering, the renormalization scale is the
virtuality of the exchanged photon, spacelike momentum transfer
squared $\mu^2 = q^2 = t$. Thus \be \alpha(t) = {\alpha(t_0) \over
1 - \Pi(t,t_0)} \ee where \be \Pi(t,t_0) = {\Pi(t) -\Pi(t_0)\over
1-\Pi(t_0) } \ee sums {\bf  all}  vacuum polarization
contributions to the dressed photon propagator, both proper and
improper. (Here $\Pi(t) =\Pi(t,0)$ is the sum of proper vacuum
polarization insertions, subtracted at $t=0$). Formally,  one can
choose any initial renormalization scale $\mu^2_0 =t_0$, since the
final result when summed to all orders will be independent of
$t_0$.  This is the invariance principle used to derive
renormalization group results such as the Callan-Symanzik
equations~\cite{Callan:1970yg,Symanzik:1970rt}.  However, the
formal invariance of physical results under changes in $t_0$ does
not imply that there is no optimal scale.  In fact, as seen in
QED, the  scale choice $\mu^2 = q^2$, the photon virtuality,
immediately sums all vacuum polarization contributions to all
orders exactly in the conventional Gell-Mann-Low scheme.   With any other choice of scale, one will recover
the same result, but only after summing an infinite number of
vacuum polarization corrections.

Thus, although the {\it initial} choice of renormalization scale
$t_0$ is arbitrary, the {\it final} scale $t$ which  sums the
vacuum polarization corrections is unique and unambiguous. The
resulting perturbative series is identical to the conformal series
with zero $\beta$-function. In the case of muonic atoms, the
modified muon-nucleus Coulomb potential is precisely
$-Z\alpha(-{\vec q}^{~2})/ {\vec q}^{~2};  $  i.e., $\mu^2=-{\vec
q}^2.$ Again, the renormalization scale is unique.


One can employ other renormalization schemes in QED, such as the
$\overline {MS}$ scheme,  but the physical result will be the same
once one allows for the relative displacement of the scales of
each scheme. For example, one can start with the result in the
$\overline {MS}$ scheme for spacelike argument $q^2=-Q^2$, for the
standard one-loop charged lepton pair vacuum polarization
contribution to the photon propagator using dimensional
regularization: \be \log {\mu^2_{\overline {MS}} \over m^2_\ell} =
6  \int^1_0 dx\, x(1-x) \log {m^2_\ell + Q^2 x(1-x)\over
m^2_\ell}, \ee  which becomes at large $Q^2$ \be
\log{\mu^2_{\overline {MS}}\over m^2_\ell} = \log{Q^2\over
m^2_\ell} - 5/3; \ee i.e., $ \mu^2_{\overline {MS}} = Q^2
e^{-5/3}.$ Thus if $Q^2 >> 4 m^2_\ell$,
we can identify \be \alpha_{\overline {MS}}(e^{-5/3} q^2) =
\alpha_{GM-L}(q^2). \ee The $e^{-5/3}$ displacement of
renormalization scales between the ${\overline {MS}}$ and
Gell-Mann--Low schemes is a result of the
convention~\cite{Bardeen:1978yd} which was chosen to define the
minimal dimensional regularization scheme.  One can use another
definition of the renormalization scheme, but the final physical
prediction cannot depend on the convention.  This invariance under
choice of scheme is a consequence of the transitivity property of
the renormalization
group~\cite{Stueckelberg,Bogolyubov:1956gh,Shirkov:1999hj,GellMann:1954fq}.

The same principle underlying renormalization scale-setting in QED
must also hold in QCD since the $n_f$ terms in the QCD $\beta$
function have the same role as the lepton $N_\ell$ vacuum
polarization contributions in QED.   QCD and QED share the same
Yang-Mills Lagrangian. In fact, one can show~\cite{Brodsky:1997jk}
that QCD analytically continues as a function of $N_C$ to Abelian
theory when $N_C \to 0$ at fixed $\alpha = C_F \alpha_s$ with $C_F
= {N^2_C-1\over 2 N_C}.$  For example, at lowest order
$\beta^{QCD}_0 ={1\over 4 \pi} \left( {11\over 3} N_C - {2\over 3
} n_f \right) \to - {1\over 4 \pi} {2\over 3 }n_f   $ at $N_C=0.$
Thus the same scale-setting procedure must be applicable to all
renormalizable gauge theories.

Thus there is a close
correspondence between the QCD renormalization scale and that of
the analogous QED process. For example, in the case of
$e^+ e^-$ annihilation to three jets, the PMC/BLM scale is set by
the gluon jet virtuality, just as in the corresponding QED
reaction.   The specific argument of the running
coupling  depends on the renormalization scheme because of their
intrinsic definitions; however, the actual numerical prediction is
scheme-independent.

The basic procedure for
PMC/BLM scale setting is to shift the renormalization scale so
that all terms involving the $\beta$ function are absorbed into
the running coupling.  The remaining series is then identical with
a conformal theory with $\beta=0.$
Thus, an important
feature of the PMC is that its QCD predictions are independent of
the choice of renormalization scheme. The PMC procedure also
agrees with QED in the $N_C \to 0$ limit.

The determination of the PMC-scale for exclusive processes is often straightforward.
For example, consider the process $e^+ e^- \to c \bar c \to c \bar
c g^* \to c \bar c b \bar b,$ where all the flavors and momenta of
the final-state quarks are identified. The $n_f$
terms at NLO come from the quark loop in the gluon
propagator. Thus the PMC scale for the differential cross section in the $\overline{MS}$ scheme
is  given simply by the $\overline{MS}$ scheme displacement of the
gluon virtuality: $\mu^2_{PMC}=e^{-5/3} (p_b+ p_{\bar b})^2$.

In practice,  one can identify the PMC/BLM scale for QCD by
varying the initial renormalization scale $\mu^2_0$ to identify
all of the $\beta$-dependent nonconformal contributions.  At
lowest order $\beta_0 = {1\over 4 \pi } \left( 11/3 N_C- 2/3
n_f\right)$. Thus at NLO one can simply use the dependence on the
number of flavors $n_f$ which arises from the quark loops
associated with ultraviolet renormalization as a marker for
$\beta_0$.

In QCD, the $n_f$ terms also arise from the renormalization of the
three-gluon and four-gluon vertices as well as from gluon
wavefunction renormalization.

It is often stated that the argument of the coupling in a
renormalization scheme based  on dimensional regularization has no
physical meaning since the  scale $\mu$ was originally introduced
as a mass parameter in extended space-time dimensions. However,
the QED example above shows that the  ${\overline {MS}} $ scale is
unambiguously related to invariants in physical 3+1 space.  The
connection of $\alpha_{\overline {MS}}$  to the Gell-Mann--Low
scheme can be established at all orders.  This also provides the
analytic extension~\cite{Brodsky:1998mf} of the
$\alpha_{\overline{MS}}$ scheme for finite fermion masses as well
to timelike arguments where the coupling is complex.

An example which shows how critical is to properly fix the
renormalization scale is the three-gluon vertex. The PMC/BLM scale
which appears in the three-gluon vertex is a function of the
virtuality of the three external gluons $q^2_1, q^2_2,$ and
$q^2_3$.  It has been computed in detail in refs.
~\cite{Binger:2006sj}. The results are surprising when the
virtualities  are very different as in the subprocess $g g \to g
\to Q \bar Q.$ \be \label{3q} \hat \mu^2 \propto {q^2_{\rm min}
q^2_{\rm med} \over q^2_{\rm max}}\ee where $|q^2_{\rm min}| <
|q^2_{\rm med}| <  |q^2_{\rm max}|$; i.e. $q^2_{\rm max}$ has the
{\bf maximal} virtuality~\cite{Lu:1992eq}. The prediction based on
simply guessing $\mu^2 \simeq q^2_{max}$ would give misleading
results.

The PMC/BLM scale that appears in the three-gluon vertex is the
mass scale that controls the number of quark flavors $n_f$ which
appears in the triangle graph. This is verified by keeping the
quark masses and threshold dynamics in the loop. Thus we
accurately determine the number of flavors $n_f$ that appears in
the $\beta$ function in the three-gluon coupling.
This generalizes for all gluonic processes.

Although these results have been obtained using the pinch-scheme,
the final PMC/BLM result is scheme-independent. The pinch scheme
is used because it provides a gauge-invariant setting for the
analysis. In effect one calculates a scattering amplitude with
three on-shell quark currents. One then obtains 14 invariant
amplitudes which describe the three-gluon vertex, only one of
which is renormalized.

In fact the calculation of the PMC scale for the three-gluon
vertex $g_a \to g_b g_c$  given in Eq.(\ref{3q}) uses the pinch
scheme to obtain a gauge invariant result.  In effect, one
computes the entire gauge invariant on-shell amplitude $q_a + \bar
q_a \to  q_b \bar q_b +q_c \bar q_c$ including the triangle loop
graph from quark loops with general mass. All 14 invariant
amplitudes are computed analytically to one loop, only one of
which is renormalized. The PMC scale for the three-gluon vertex as
given in Eq.(\ref{3q}) also correctly sets the scale which
controls the number of effective flavors which contribute to the
$\beta-$ function for the three-gluon vertex. Details are given in
refs \cite{Binger:2006sj},\cite{Lu:1992eq}.

These results show that the usual method of guessing the
renormalization scale for processes involving the three-gluon and
four-gluon couplings, typically misses this essential physics,
assigns $n_F$ incorrectly and mischaracterizes the perturbative
prediction.  The error which is introduced can be in principle
eliminated at infinite order, but only if one can sum the
renormalon series.

 The explicit result for the PMC/BLM scale
is the physical scale controlling the quark threshold in the
specific renormalization procedure used, but it is always possible
to relate one scheme with another by the transitivity property of
the renormalization group. This property is guaranteed by the PMC
so there can be a constant displacement between schemes.

The PMC method is a general approach to set the
renormalization scale in QCD including purely gluonic processes. It is scheme independent and void of
renormalon growth due to the absence of the $\beta-$ function terms in the
perturbative expansion.
We stress that the $\beta$-function is gauge invariant in any
correct renormalization scheme. The resulting conformal series is then
gauge invariant. Thus the PMC is a gauge-invariant procedure.

It is sometimes argued that it is advantageous not to fix the
renormalization scale at all, since its variation provides a
measure of higher-order contributions to the theory predictions.
In fact, one obtains sensitivity only to the $\beta$-dependent
non-conformal terms by this procedure.  In some cases the
conformal contributions may be unexpectedly large. For example,
the very large electron-loop light-by-light scattering
contribution~\cite{Aldins:1970id} $\simeq 18 (\alpha^3/\pi)^3 $
to the muon anomalous magnetic moment is unassociated with
renormalization or the $\beta$ function. Of course, one can still
compute the variation of the prediction around the PMC scale as an
indicator of higher order non-conformal terms.

Stevenson has proposed that one should set the renormalization
scale at a point where the predicted cross section has minimal
variation with respect to $\mu$ -- the ``principle of minimal
sensitivity" (PMS) ~\cite{Stevenson:1981vj}.  However, unlike the
PMC, the application of the PMS to jet  production gives
unphysical results~\cite{Kramer:1987dd} since it  sums physics
into the running coupling not associated with renormalization.
Worse, the PMS prediction depends on the choice of renormalization
scheme, and it violates the transitivity property of the
renormalization group~\cite{Brodsky:1992pq}.  Such heuristic
scale-setting methods also give incorrect results when applied to
Abelian QED.

It should be emphasized that the {\it factorization scale}  which
enters predictions for  QCD inclusive reactions is introduced to
match nonperturbative and perturbative aspects of the parton
distributions in hadrons; it is present even in conformal theory,
and thus its determination is a completely separate issue from
{\it renormalization scale} setting.

\section{Identifying the Renormalization Scale using the Principle of Maximum Conformality}

Given the analytic form of the hard process amplitude or cross
section as a series in $\alpha_s(\mu^2_0)$ calculated at an
initial scale $\mu^2_0$  and at a certain order (NLO, NNLO and so
on) , one can identify the PMC scale, order by order, in a
systematic way:

\begin{enumerate}

\item The variation of the cross section with respect to $\log
\mu^2_0$ can be used to distinguish the conformal terms versus the
nonconformal terms proportional to the $\beta$ function.

\item The identified nonconformal terms have the form $\beta
\times \log{p_{ij}/ \mu^2_0}$ where $p_{ij} = p_i \cdot p_j$  are
the scalar product invariants $i \ne j$ which enter the hard
subprocess.  In practice, these terms can be identified as
coefficients of  $n_f$, the number of flavors appearing in the
$\beta$ function; i.e., the flavor dependence arising from quark
loops associated with coupling constant renormalization.  The
$n_f$ terms in QCD arise from the renormalization if the
three-gluon and four-gluon vertices as well as from gluon
wavefunction renormalization.

\item The scale is then shifted $\mu^2_0 \to  \mu^2$ in order to
absorb the non-conformal terms. Thus when the scale is correctly
set, the coefficients of $\alpha_s(\mu^2)$ become independent of
the $\beta$ function and $\log { \mu^2}. $

\item  The series is then identical to that  of the conformal theory where $\beta=0$ as given by the
Banks-Zaks method~\cite{Banks:1981nn}.

\item The PMC scale is fixed for an observable (such as a
differential cross section). PMC then can give a single effective global scale for
the whole set of skeleton graphs entering the calculations which
sums all the non-conformal $\beta$-terms associated with renormalization into the running
coupling.

Other examples of this procedure will be given in the next
sections.
\end{enumerate}

\subsection{The Global PMC Scale}
Ideally, as in the BLM method, one should allow for separate
scales for each skeleton graph; e.g., for  electron-electron
scattering, one takes $\alpha(t)$ and $\alpha(u)$ for the
$t$-channel and $u$-channel amplitudes, respectively.

Setting separate renormalization scales can be a challenging task
for complicated processes in QCD where there are many final-state
particles and thus many possible Lorentz scalars $p^2_{ij} =
p_i\cdot p_j.$  However, one can obtain a useful first
approximation to the full PMC/BLM scale-setting procedure by using
a single {\it global} scale  $\mu^2$ which appropriately weights
the individual BLM scales.

The global scale can be determined by  varying the subprocess
amplitude with respect to each invariant, thus determining the
coefficients $f_{ij}$ of $\log {p^2_{ij} /\mu^2_0}$ in the
nonconformal terms in the amplitude. The global PMC scale is then
\be { \mu}^2 = C \times  \Pi_{ij} ~ [p^2_{ij}]^{w_{ij}}, \ee i.e.,
\be \log { \mu^2} = \sum_{i\ne j} w_{ij} \log{p^2_{ij}} + \log C
\ee where the weight for each invariant is \be w_{ij} = { f_{ij}
\over \sum_{i\ne j} f_{ij}}. \ee and $\sum_{i\ne j} w_{ij} =1.$
The constant $C$ is the scheme displacement; e.g., $C= e^{-5/3}$
for $\overline {MS}$ for $ \mu^2 >> 4m^2_f.$

As a specific example of the application of  a  PMC global scale,
consider the electron-electron scattering amplitude in QED. (For
simplicity,  we will just take the contribution of the convection
current to the amplitude, as in scalar QED.) The Lorentz invariant
Born amplitude at the initial scale $t_0$ is then \be M^0(t,u) = 4
\pi \alpha(t_0) \big(\frac {s-u}{t} + \frac{s-t}{u}\big) . \ee The
running QED coupling $\alpha(q^2)$  in QED sums all proper and
improper vacuum polarization graphs \be M(t,u) = 4 \pi \alpha(t)
\big(\frac{s-u}{t}\big) + 4\pi \alpha(u) \big(\frac{s-t}{u}\big)
 \ee
where to leading order \be \alpha(t)=\alpha(t_0)\big(1+n_\ell\,
\frac{\alpha(t_0)}{3 \pi} \, \log{-t \over t_0}\big). \ee Aside
from power-suppressed contributions involving the lepton masses,
the resulting series is identical to the corresponding conformal
theory with $\beta=0.$

In this process we have contributions from both the $t$ - and $u$-
channel amplitudes which require separate renormalization scales
for each skeleton graph. However, at leading order we can weight
the amplitudes to obtain a single PMC/BLM scale which still sums
the nonconformal $\beta$ terms into the running coupling
$\alpha(\mu^2)$ at leading order. For example, using the standard
Gell-Mann--Low scheme, we can write \be M(t,u)= f(t) \alpha(t)+
g(u) \alpha(u)= (f(t)+g(u))\alpha(\hat \mu^2) \ee where $f(t)=
4\pi (s-u)/t$ and $g(u)=4\pi (s-t)/u$ are the Born amplitudes for
the $t$ - and $u$ -channels, respectively.

Then in this case we have two basic PMC scales $\alpha(t)$  and
$\alpha(u)$ for each skeleton graph in the standard Gell Mann-Low
scheme used in QED. These couplings then sum all of the vacuum
polarization corrections to the skeleton graphs to infinite order.
The result is then gauge invariant and the logarithm of the global
scale is \be \label{eq14} \log{ \hat \mu^2 }= {f(t)\over f(t)+
g(u)} \log{(- t)}  + {g(u)\over f(t)+ g(u)} \log{( - u)} \ee

which duplicates the multi-scale result at NLO.

One can also use the mean value theorem to obtain an effective
single scale which analytically reproduces the exact multi-scale
result to next to leading order. Since it matches the exact result
at NLO, it also retains gauge invariance at this order.  Moreover,
the PMC single or multi-scale result is independent of the choice
of scheme. The single scale result illustrates why it is wrong to
guess a single scale like  $\mu^2 = p^2_T$ since it fails to agree
with this simple example.

 Using kinematical constraints such as the total momentum conservation $s+t+u=0$
 the weighted scale dependence can be confined into the $\log(t/u)$
term inside the running coupling. The global scale $\hat \mu^2$ is
maximal at $\theta_{CM} =\pi/2$ ($\mu^2 = \sqrt{tu}= -t= -u$) and
vanishes at the boundaries $(0,\pi)$ where $
\tan^2({\theta_{CM}/2})= t/u.$
The effective renormalization scale for electron-electron
scattering in Eq. \ref{eq14} is weighted by the respective
scattering amplitudes.  The $t$-channel amplitude strongly
dominates at $\Theta_{CM}=0$, and the renormalization scale is
thus $t$. Similarly, the $u$ - channel amplitude strongly
dominates at $\Theta_{CM}=\pi$,  and the effective renormalization
scale  in that domain is $u$.  Thus in both limits the effective
renormalization scale $ \hat \mu$ vanishes.

\begin{figure}[hbpt]
\begin{center}
   \includegraphics[width=10cm]{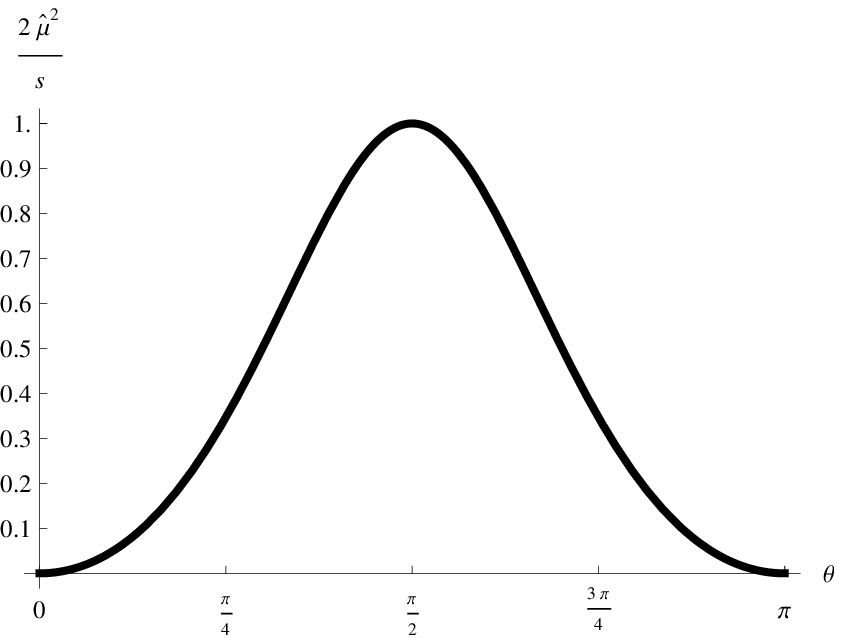}\\
  \caption{The PMC/BLM scale as function of the CM angle $\theta_{CM}$:$e \, e
\rightarrow e \, e$ scalar QED}\label{pmcnorm}
  \end{center}
\end{figure}

The results are shown in Fig. \ref{pmcnorm}.

\section{A PMC Example for QCD: Application to Jet Cross Sections in Electron-Positron Annihilation}

As an example of the application of the PMC to QCD, we will show
how the renormalization scale can be determined for the cross
sections for $e^+ e^-$ annihilation into two and three jets in
$\overline{MS}$ scheme.

The two-jet cross section has only infrared divergences:
\be
\sigma^{(2)} =  \sigma_0 \big({4\pi \mu^2\over q^2} \big)^{\lambda/2}  \big(1-\lambda/2\big)
{\Gamma(1-\lambda/2)\over\Gamma(2-\lambda)}
\ee
where $\sigma_0 = 4\pi {\alpha^2\over 3 q^2}N_C\sum^{N_f}_{i=1} e^2_i.$

Here $\lambda \equiv 4- n $ is the number of extra space-time
dimensions used to regulate infrared and ultraviolet divergent
integrals. Eventually all of the infrared divergences and the factors
involving $\lambda$ will cancel out.   In dimensional regularization
the scale $\mu$ is introduced as a mass scale to restore the
correct dimension of the coupling. The gauge coupling $g_R$ is
related to the renormalized coupling constant $\alpha_R$ by
\be\frac{g_R^2}{(4 \pi)^{(4-\lambda)/2}}=\frac{\alpha_s(\mu^2)}{4
\pi} (\mu^2)^{\lambda/2} e^{\gamma_E \lambda/2}\ee and here
$\gamma_E$ is the Euler constant.

As discussed in the
introduction, the mass scale of schemes defined by dimensional
regularization attains its physical meaning  when it is applied to
QED. The renormalized gauge coupling is also related to the bare
coupling by:
\be
g_R=\sqrt{Z_3} Z_2 /Z_1 g_0,
\ee
where $Z_1$ is
the renormalization constant for the quark-antiquark-gluon vertex,
$Z_2$ for the quark field and $Z_3$ for the gluon field. The
renormalization constants are:
\bea
Z_1&=& 1-\frac{g^2_0}{16
\pi^2}\,\, (N_c+C_F)\, \left(
\frac{2}{\lambda_{UV}}-\frac{2}{\lambda_{IR}}\right) \\
Z_2&=& 1-\frac{g^2_0}{16 \pi^2}\,\, C_F \, \left(
\frac{2}{\lambda_{UV}}-\frac{2}{\lambda_{IR}}\right) \\
Z_3 &=& 1+\frac{g^2_0}{16
\pi^2}\,\,(\frac{5}{3}N_c-\frac{2}{3}N_f)\, \left(
\frac{2}{\lambda_{UV}}-\frac{2}{\lambda_{IR}}\right)
 \eea
 where $\lambda_{UV},\,\lambda_{IR}$ are related respectively to the $UV-$ultraviolet and $IR-$infrared poles.
 In the ${MS}$ only the pole associated with UV renormalization is subtracted out, and this leads us
 to a redefinition of the gauge coupling:
\be
 \frac{1}{g_R} \, \delta g_0 =
\frac{g_R^2}{16\pi^2}(\frac{2}{3}N_f -\frac{11}{3}N_c) \,
\frac{1}{\lambda_{UV}} \ee A suitable renormalization scheme is
the $\overline{MS}$ which differs from $MS$ by a constant term and
the respective counterterm can be inserted in the Born cross
section by shifting the coupling constant:
 \be \label{msbar}
\alpha_s^0=\alpha_s^{\overline{MS}} \, \left\{1- \left(
\frac{11}{6}N_c -\frac{2}{3}T_R \right)
\frac{\alpha_s^{\overline{MS}}}{2\pi} \,\left( \frac{1}{\epsilon}
+{(\ln{4\pi}-\gamma_E)} \right) \right\}= \alpha_s^{\overline{MS}}
\, \left\{1- \, \beta_0 \alpha_s^{\overline{MS}} \,\left(
\frac{1}{\overline{\epsilon}}\right) \right\}
 \ee
 where:
 \be
 \label{epsilonbar}
 {1 \over {\overline \epsilon}} ={1 \over {
 \epsilon}}+(\ln{4\pi}-\gamma_E),
 \ee
\be \beta_0= \frac{1}{2\pi} \left( \frac{11}{6}N_c -\frac{2}{3}T_R
\right)
 \ee
 with $T_R=N_f/2\,,\, \epsilon=\lambda_{UV}/2.$ \\

The Born cross section for $e^+ e^- \to q(p_1) \bar q(p_2) g(p_3)$
for massless quarks and gluons is \be \label{born3j} \left.
{d\sigma^{(3)} (\mu^2) \over dx_1 dx_2}\right|_{\rm Born} =
\sigma^{(2)} \,\, \big( {4\pi \mu^2\over q^2} \big)^{\lambda/2}
\,\, {1\over\Gamma(1-\lambda/2)} \, F_\lambda(x_1,x_2) \, {
\alpha_s^{MS}(\mu^2) \over 2\pi} \,\, C_F B^{V-\lambda/2
S}(x_1,x_2) \ee Here \be F_\lambda (x_1,x_2) = [(x_1+x_2 -1)
(1-x_1)(1-x_2)]^{-\lambda/2} \ee and \be B^{V-\lambda/2
S}(x_1,x_2) = B^V(x_1,x_2) - {\lambda\over 2}B^S(x_1,x_2) \ee \be
B^V(x_1,x_2) = {x^2_1+ x^2_2\over (1-x_1)(1-x_2)} \ee \be
B^S(x_1,x_2)  = {x_3^2\over (1-x_1)(1-x_2)} \ee where $x_i = {2E_i
\over \sqrt {q^2}}$ in the $e^+ e^-$ CM. In terms of invariants:
$y_{ij} = s_{ij}/q^2 = (p_i+p_j)^2/q^2$. Then
 $x_1 = 1-y_{23}, x_2 = 1-y_{13}, x_3 = 1-y_{12}, x_1+x_2+x_3 = 2$.

The renormalized one-loop corrected cross section for $e^+ e^- \to
q(p_1) \bar q(p_2) g(p_3) $ is given by Eq. (2.11) of Fabricius et
al.~\cite{Fabricius:1981sx}
For our purposes it is sufficient to
quote only the term proportional to  $\beta_0$ in the
$\overline{MS}-$scheme:
\be
\left. {d\sigma^{(3)}\over dx_1 dx_2}\right|_{\rm one loop}
=\left. {d\sigma^{(3)}(\mu^2)\over dx_1 dx_2}\right| _{\rm
Born}\,\, \left[1+ { \alpha_s(\mu^2)}
 {\Gamma(1-\lambda/2)\over\Gamma(1-\lambda)}\big({4\pi \mu^2\over
 q^2}\big)^{\lambda/2}\,\,
{ \beta_0}\big( \log{\mu^2\over q^2}\big) + \cdots\right] \ee
where the coupling is defined as in Eq. \ref{msbar}:
$\alpha_{MS}(e^{ \log{4\pi} -\gamma_E} \mu^2) \equiv
\alpha_{\overline {MS}}(\mu^2).$ The remaining contributions are
independent of $n_f$ and $\beta_0$

We can eliminate the non-conformal log-term proportional to $\beta_0$ by shifting the renormalization scale
$\alpha_{MS}(\mu^2) $ in the Born cross section Eq. \ref{born3j}
$$
\alpha_s(\mu^2)\, \simeq \, \alpha_s(q^2)\left(1-\alpha_s(q^2)\,
\beta_0\, \log[{\mu^2 \over q^2}]\right);$$
however, it is first convenient to shift the
scale to $\mu^2 \to ( \mu^2_0) .$

Then
\be
\left. {d\sigma^{(3)}\over dx_1 dx_2}\right|_{\rm one
loop} = \left. {d\sigma^{(3)}(\mu_0^2)\over dx_1 dx_2}\right|
_{\rm Born}\,\, \left[1+ { \alpha_s(\mu^2_0)}
 {\Gamma(1-\lambda/2)\over\Gamma(1-\lambda)}\big({4\pi \mu_0^2\over
 q^2}\big)^{\lambda/2}\,\,
{ \beta_0}\big( \log{\mu_0^2\over q^2}\big) + \cdots\right]
\ee

Naively one could simply fix the scale to  $\sqrt{q^2}$, but
the 3-jet cross section will still be affected by  IR divergences;
in order to apply the PMC/BLM prescription we will first need  to
include the 4-jet contributions.

\section{Numerical scale fixing}
The complete differential 3-jet cross section has been calculated
by Fabricius et al.~\cite{Fabricius:1981sx}, and we quote here the
results for the $\beta_0- $dependent terms:

\be
{d^2\sigma^{(3)}(\epsilon,\delta)\over dx_1 dx_2} = \sigma_0
{\alpha_s(q^2)\over 2 \pi} C_F\,\, \times
\ee
\be \left\{
B^V(x_1,x_2) \left[1-{\alpha_s(q^2)}\,\, \beta_0 \,\,
\left(\log({1-\,\, \cos\delta\over 2})+ \,\, \log \hat{x}^2_3
-\,\, \frac{13}{3}\right)\right]
 -B^S(x_1,x_2)\, {\alpha_s(q^2)} \, {\beta_0 \over 2} \right\} +
{\cal{O}}(\delta^2)) + \cdots
 \ee
 where $\hat{x}_3=(2-x_1-x_2)$
and \be
d\sigma^{(3)}(\epsilon,\delta)=d\sigma^{(3)}+d\sigma^{(4)}(\epsilon,\delta)
\ee is the sum of the 3- and the 4-jets contributions.  The
cancellation of the IR-poles is guaranteed by the KLN
theorem~\cite{Kinoshita:1962ur,Lee:1964is}.

The variables $(\epsilon,\delta)$ are small quantities introduced
in the virtual amplitude in order to define the soft and collinear
4-jet contributions to the 3-jet cross section. In particular
these quantities refer respectively to the fraction of the total
energy and to the cone opening angle which define the phase volume
for a 3-jet event (for more details, see
Ref.~\cite{Fabricius:1981sx}).

 In order to extract the PMC/BLM scale we  first work in the $\overline{MS}$-scheme,
 fixing an arbitrary renormalization scale: $\mu^2=\mu^2_0.$
It turns out that $\beta_0$ term of the 3-jet differential IR safe
cross section has the form:
 \bea
 {d^2\sigma^{(3)}(\epsilon,\delta)\over dx_1 dx_2} &=&
\sigma_0 {\alpha_s(\mu_0^2)\over 2 \pi} C_F \times \\
& &  \left\{ B^V(x_1,x_2) \left[1-{\alpha_s(\mu^2_0)} \, \beta_0
\,\, \left(\log({1-\cos\delta\over 2})+ 2 \, \log{ (2-x_1-x_2)} -
{13\over 3} +\log{q^2 \over \mu_0^2 } \right)\right]\right.   \nonumber \\
& & \left. - B^S(x_1,x_2) {\alpha_s(\mu^2_0)} {\beta_0 \over 2}
\right\} + {\cal{O}}(\delta^2)) + \cdots. \nonumber \eea In
principle we can extract information on the terms in this formula
performing a detailed analysis of the dependence of the
$\beta_0-$coefficient on the invariants. Performing a blindfold
study we can single out the $\beta_0-$coefficient by means of the
$\beta_0-$derivative  of the whole cross section or either by the
$n_f-$derivative since:

\be
\label{betanf} \frac{d
f}{d\beta_0}=\frac{d f}{dn_f} \times \frac{d \beta_0}{dn_f}^{-1}
\ee

Then we can factorize out the Born amplitude Eq.\ref{born3j}: \bea
\left. {d\sigma^{(3)}(\mu_0^2)\over dx_1 dx_2}\right|^{-1}_{\rm
Born} \,\cdot\, {d \over d
\beta_0}{d^2\sigma^{(3)}(\epsilon,\delta;\mu_0^2)\over dx_1 dx_2}
&=& \left[ -{\alpha_s(\mu^2_0)}  \,\, \left(
\log({1-\cos\delta\over 2})+ 2 \, \log{ (2-x_1-x_2)} - {13\over 3}
+\log{q^2 \over \mu_0^2 }  \right. \right. \nonumber  \\ & &
\left. \left.+
 {B^S(x_1,x_2)\over 2 \,
B^V(x_1,x_2)}\right)\right]+{\cal{O}}(\delta^2)) + \cdots.
\nonumber \eea and at the first order approximation the PMC/BLM
scale can be fixed numerically imposing: \be \left. \left[ \left.
{d\sigma^{(3)}(\mu^2)\over dx_1 dx_2}\right|^{-1}_{\rm Born}
\,\cdot\, \left(\left. {d \over d
n_f}{d^2\sigma^{(3)}(\epsilon,\delta;\,\mu^2)\over dx_1
dx_2}\right)\right|_{n_f=0} \right]\right|_{\mu^2={\mu}^2_{PMC}} =
\,\, 0 \ee

In the numerical procedure at NLO the analytic form of the cross
section is not needed; one must only keep track of the appearance
of number of flavors $n_f$ arising from loop diagrams involving
renormalization. This procedure ,which has been shown at NLO here,
can also be iterated to higher orders in $ \alpha_s$, by keeping
track of the $n_f$-terms entering the $\beta$-function, leading us
to an improvement of the accuracy of the PMC/BLM scale
${\mu}^2_{PMC}.$

Following this procedure we can include all the non-conformal
$\beta$ terms into the running coupling constant for every
physical process, setting the renormalization scale at the PMC/BLM
scale without necessarily knowing the PMC/BLM analytic form. Thus
we end up with a cross section which is formally equal to the
corresponding conformal expansion with $\beta=0.$ In this
particular case the PMC/BLM scale has the form: \be
\mu^2_{PMC}\simeq q^2\,\,(2-x_1-x_2)^2 \,\, {\delta^2 \over 4}\,
e^{-\frac{13}{3}+\frac{B_S(x_1,x_2)}{2 \, B_V(x_1,x_2)}}.\ee In
this case the coefficient depends on the parton energies $x_1,
x_2,$ on the angle parameter $ \delta,$ and on the scale ratio
$q^2/\mu_0^2$ (all these quantities can be written in the form of
Lorentz invariants). The different contributions to the
coefficient can be also identified, term by term, by considering
the most differential cross section (i.e. for the 3-jet case the
triple differential cross section),  by performing the derivative
(or logarithmic derivative) with respect to the corresponding
invariant, and then isolating the constant term. This procedure
will be discussed in detail in the next section.

\section{The PMC/BLM scale as a function of the jet mass resolution parameter}

As shown by Kramer and Lampe \cite{Kramer:1987dd}, one can define
a QCD jet by defining a resolution parameter $y\cdot s$ as its
maximal virtuality. The jet then consists of particles with total
invariant mass squared smaller than $y \cdot s$.  Using this
definition, we will perform the integration of the entire
three-jet differential cross section, including real,
$d\sigma^{(3)} $, and virtual, $d\sigma^{(s)} $, contributions in
order to have a IR safe quantity. This gives a $y$-dependent
integrated formula with $\beta_0$ dependent terms which can be
absorbed into the argument of the running coupling, according to
the PMC/BLM prescription.

The entire differential three-jet cross section \cite{ellis1981}:
\be \frac{1}{\overline \sigma_0} \,
\frac{d\sigma^{(s)}+d\sigma^{(3)}}{dy} = \int_{y}^{1-2 y} dz
\int_{y}^{1-y-z} dx \,\,T[1-x-z,x,z] \alpha_s(Q^2) (1 - \beta_0
\,\alpha_s (Q^2) (\log[x]+\log[z]-\frac{5}{3}......))
 \nonumber \ee \be
 = \alpha_s(Q^2)\, (T(y) - \beta_0\, \alpha_s(Q^2) \,( C(y)+....))
 \ee

 \be
 \equiv  T(y)
 \alpha_s(Q^2) \,(1 -\beta_0 \, \alpha_s(Q^2) \, 2\log[\frac{\mu_{BLM}}{\sqrt{s}}])=
 T(y)\alpha_s(\mu^2_{BLM});
 \ee
where : ${\overline \sigma_0}={\sigma_0} \, C_F \, Q^2 \, /{2\pi }
\,,s=Q^2\,,x=y_{13}\,,z=y_{23},$ \be T[x_1,x_2,x_3]\, =\, \frac{2
x_1^2+x_2^2+x_3^2+2 x_1 (x_2 +x_3 )}{x_2  x_3 } \ee

and $T(y),C(y)$ result from the partial integration of the LO- and
NLO- terms of the 3-jet cross section (for more details see Ref.
\cite{ellis1981}\cite{Kramer:1987dd}).

\begin{figure}[hbpt]
\begin{center}
  \includegraphics[width=10cm]{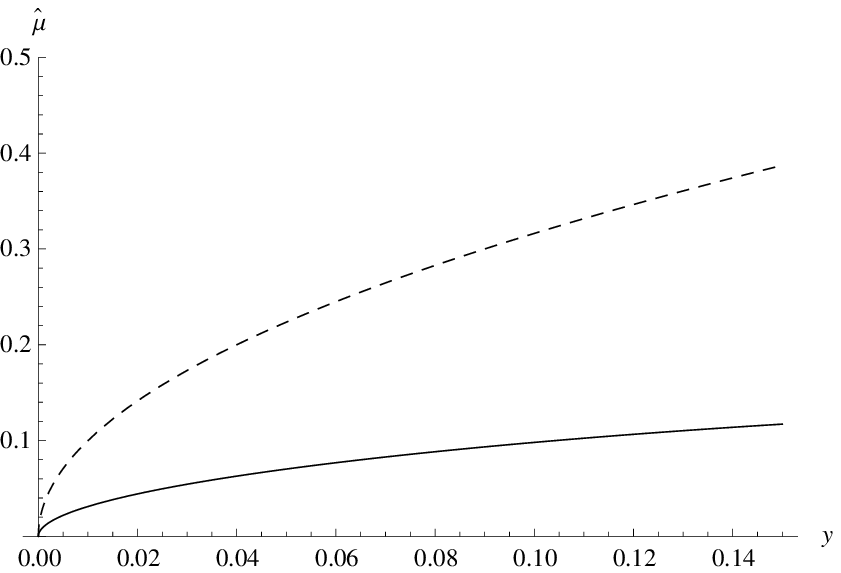}\\
  \caption{The PMC/BLM scale, $\mu_{PMC}$ (plane line) as a function of the jet resolution parameter $\textit{y},$
  for $e^+ e^- \to q \bar q g$.}\label{kramer}
 For comparison, the behavior $\hat{\mu}\simeq \sqrt{y}$ is also shown (dashed line).
  \end{center}
\end{figure}

Then in the 3-jet case, the BLM-PMC scale as function of the
jet-virtuality $y$, has the analytic form: \be {\hat \mu^2} =
\mu^2_{PMC/BLM}= s \, \times \, e^{-\frac{5}{3} \,+\,
\frac{C(y)}{T(y)}} \ee A plot of the PMC/BLM scale against $y$,
the virtuality resolution of the jet, in $e^+ e^- \to q \bar q g$
is shown in Fig. \ref{kramer}. The result agrees with the BLM
scale calculated by Kramer and Lampe in the $\overline{MS}$
scheme. The PMC/BLM prediction is scheme-independent; the specific
value of the renormalization scale is rescaled according to  the
choice of scheme so that all results are commensurate. The PMC/BLM
scale also accurately determines $n_f$, the effective number of
flavors in the $\beta$-function. As is clear from the QED analog,
the renormalization scale reflects the virtuality of the gluon
jet; it thus must vanish when the resolution $y\,s$ vanishes.  As
noted by Kramer and Lampe~\cite{Kramer:1987dd}, the
renormalization scales determined by the {\it ad hoc} PMS and FAC
(Fastest Apparent Convergence)~\cite{Kubo:1982gd} procedures have
the wrong physical behavior at $y\,s \to 0,$ since they become
infinite $\mu^2 \to \infty$ as the jet resolution and gluon
virtuality vanish.

\section{PMC/BLM scale fixing in the 3 Jet case:  The complete differential cross section}

In the case of the complete differential cross section; i.e., the
most differential cross section for a given process  without any
constrained variables, the PMC/BLM scales depend on the number of
flavors $n_f$ and on the independent invariants entering the
process. In the case of the three jets, we notice that the cross
section depends on the color and flavor parameters $n_f,N_C,C_F$
and on the kinematical invariants $s_{12},s_{13},s_{23}$ where the
label $3$ refers to the gluon momentum, and the indices $1,2$
refer to the quark and anti-quark momenta. On the other hand, the
nonconformal terms entering the running coupling depend only on
the number of flavors $n_f$ and on a reduced number of kinematical
invariants. These terms can be identified by first varying the
number of flavors $n_f$ and then the invariant $s_{ij}$, whereas
the constant term can be extracted by simply subtraction at the
final step.  Starting with the triple differential cross section
for three jets, which is given by the sum of the singular part of
4-jet differential cross section $d\sigma^{(s)}$ and the real
3-jet cross section $d\sigma^{(3)}$ ( for more details see
Ref.\cite{ellis1981}):

 \be \frac{d\sigma^{(s)}+d\sigma^{(3)}}{dz \, dy\, dx}=
{\tilde{\sigma}_0} \, \frac{\alpha_s(Q^2)}{2 \pi} \,
\delta(1-x-y-z)\Big\{ T[z,x,y] \,\, \Big[1+ \frac{\alpha_s(Q^2)}{2
\pi} C_F (....)+ \frac{\alpha_s(Q^2)}{2 \pi} N_C (....)  \nonumber
\ee \be  - \alpha_s(Q^2) \,\, \beta_0
\,\,\Big(\log[x*y]-\frac{5}{3}\Big)\Big]+\frac{\alpha_s(Q^2)}{2
\pi} F[z,y,x]\Big\}\ee with $\tilde{\sigma}_0=\sigma_0\, C_F s.$
For simplicity sake we are using the notation $(z,x,y)$ for
respectively the final state gluon-, quark-, antiquark-energy. In
order to extract the first order terms related to the $\beta-$
function we can start performing an \textit{ab initio} analysis of
the cross section. We can first single out the $\beta_0$
coefficient by means of the $\beta_0-$ derivative, or either by
the number of flavors $n_f-$derivative, using Eq. \ref{betanf} and
then we can factorize out the Born amplitude:
 \be \left. \frac{d\sigma^{(3)}(Q^2)}{dz \, dy\,
dx}\right|^{-1}_{Born} {1 \over \alpha_s(Q^2)} \frac{d}{d\beta_0}
\,\,  \left( \frac{d\sigma^{(s)}+d\sigma^{(3)}}{dz \, dy\,
dx}\right) =
 \, \left[\log[x\,
y]-\frac{5}{3}\right]+O(\alpha_s),
 \ee

 $$\left. \frac{d\sigma^{(3)}(Q^2)}{dz \, dy\, dx}\right|_{Born} ={\tilde{\sigma}_0} \, \frac{\alpha_s(Q^2)}{2 \pi} \, T[z,x,y] \,\,
\delta(1-x-y-z).$$ Finally, we can extract the weight for each
invariant by taking the logarithmic derivative: \be
\omega_i=\frac{d}{d\log(x_i)} \,\,  \left. \left(
\frac{d\sigma^{(3)}(Q^2)}{dz \, dy\, dx} \right|^{-1}_{Born} \,\,
{1 \over \alpha_s(Q^2)}\, \frac{d}{d\beta_0}\,\,   \left(
\frac{d\sigma^{(s)}+d\sigma^{(3)}}{dz \, dy\, dx}\right) \right)
\ee where $x_i=(x,y,z).$ The constant term can be identified by
subtracting out all the logarithm terms from the $\beta_0$
coefficient. Then at first order approximation in the coupling
constant, the $\mu_{PMC}$-scale for the 3-jet differential cross
section has the analytic form: \be \label{pmc3jetdiff}
\mu^2_{PMC}\, \simeq \,Q^2\times \,C \,\times \,\prod_i \, x_i^{\,
\omega_i} = Q^2 \,\, x\,\, y \,\, e^{-\frac{5}{3}}.
 \ee

\subsection{Commensurate Scale Relations}

Relations between observables must be independent of the choice of
scale and renormalization scheme.  Such relations, called
``Commensurate Scale Relations"(CSR)
~\cite{Brodsky:1994eh,Brodsky:1995tb,Broadhurst:2004jx} are thus
fundamental tests of theory, devoid of theoretical conventions.
One can compute each observable in any convenient renormalization
scheme, such as the $\overline{MS}$ scheme using dimensional
regularization. However, the relation between the observables
cannot depend on this choice - this is the transitivity property
of the renormalization
group~\cite{Stueckelberg,Bogolyubov:1956gh,Shirkov:1999hj,GellMann:1954fq}.
For example, the PMC relates the effective charge
$\alpha_{g_1}(Q^2)$, determined by measurements of the Bjorken sum
rule, to the effective charge $\alpha_R(s)$, measured in the total
$e^+ e^-$ annihilation cross section: $ [1-\alpha_{g_1}(Q^2)/\pi
]\times  [1+ \alpha_R(s^*)/ \pi] = 1.$ The ratio of PMC scales
$\sqrt s^*/Q \simeq 0.52$  is set by  physics;  it guarantees that
each observable goes through each quark flavor threshold
simultaneously as  $Q^2$ and $s$ are raised. Because  all $\beta
\ne 0$ nonconformal terms are absorbed into the running couplings
using PMC, one recovers the conformal
prediction~\cite{Brodsky:1995tb}; in this case, it is the Crewther
relation~\cite{Crewther:1972kn,kataevadd1,kataevadd2,kataevadd3,Kataev:2011zh}.
Thus by applying the PMC, the conformal commensurate scale
relations between observables, such as the Crewther relation,
become valid for non-conformal QCD at leading twist.

\section{Conclusions}
As we have shown,  the principle of maximal conformality (PMC)
provides a consistent method for setting the optimal
renormalization scale in pQCD.
The PMC scale is determined by
identifying the $\beta$ terms in the next-to-leading
contributions and making the appropriate shift in order to include
the $\beta$-terms into the running coupling. This can be done most
simply by identifying the $n_f$ terms which come from quark loops
of skeleton graphs. This includes the $n_f$ terms which
renormalize the three and four gluon couplings. This procedure has been
used to identify the
correct PMC scale for the three-gluon vertex \cite{Binger:2006sj}
\cite{Lu:1992eq}. The resulting series is identical to that  of
the corresponding conformal theory with $\beta=0$ as given, for
example,  by the Banks-Zaks method~\cite{Banks:1981nn}.

The global PMC renormalization scale is particularly useful for
very complex processes; one only requires the dependence of the
calculated subprocess amplitudes on the initial renormalization
scale $\mu^2_0$ and $n_f$, the number of quark flavors appearing
from quark loops associated with renormalization. The single
global PMC scale, valid at leading order, can thus be derived from
basic properties of the perturbative QCD cross section.

We have discussed specific methods for efficiently determining the
PMC renormalization scale analytically or numerically for QCD hard
subprocesses. The analytic form of the PMC renormalization scale
can be determined by varying the subprocess amplitude with respect
to each invariant, thus determining the coefficients $f_{ij}$ of
$\log {p^2_{ij} /\mu^2_0}$ in the nonconformal terms in the
amplitude. This result can be used to fix the renormalization
scales for each contributing skeleton graph. However, we have
shown that a single PMC  global scale can then determined at NLO
by appropriate weighting. Alternatively the numerical value of the
PMC scale can be determined without  specific information on the
analytic form from the $n_f$-derivative of the cross section. The
two methods give rise to the same results at NLO.

The factorization scale, in contrast, is the scale entering the
structure and fragmentation functions. Unlike the renormalization
scale, a factorization scale ambiguity occurs even in a conformal
theory. The factorization scale should be chosen to match the
nonperturbative bound state dynamics with perturbative DGLAP
evolution. This could be done explicitly using nonperturbative
models  such as AdS/QCD and light-front holography where the
light-front wavefunctions of the hadrons are known.

Note that one applies the PMC method to renormalizable hard
subprocesses (including the associated radiation diagrams required
for IR finiteness) which enter the pQCD leading-twist
factorization procedure. The initial and final quark and gluon
lines are taken to be on-shell so that the calculation of the hard
subprocess amplitude is gauge invariant. Thus the application of
the PMC to hard subprocesses does not involve the factorization
scale, and thus no double or single logarithms which involve the
factorization scale enter.

The usual heuristic method of guessing the renormalization scale
and varying it over a range of a factor of two gives
scheme-dependent results, leaves the non-convergent perturbative
series and gives the wrong result when applied to QED processes.
In fact, varying the renormalization scale around
such a guess only exposes nonconformal contributions involving the
$\beta$ function; it gives no information on the conformal
contributions. The PMS method~\cite{Stevenson:1981vj} has similar
faults -- it violates the transitivity property of the
renormalization group, depends on the choice of scheme,  is wrong
for QED, and as shown by Kramer and Lampe~ \cite{Kramer:1987dd},
leads to unphysical results.
In contrast, the PMC method, which has no such disadvantages, and
satisfies all principles of renormalization theory, gives the
optimal prediction for pQCD at each finite order.

The PMC is  the theoretical principle underlying the BLM procedure
and commensurate scale relations between observables - the
rigorous scale-fixed scheme-independent relations in QCD between
observables, such as the Generalized Crewther relation; it is also
the scale-setting method used for precision determinations of
$\alpha_s$ in lattice gauge theory~\cite{Davies:2008sw}. In
addition, it has been recently shown that for certain observables
in 2-jet production the results of the MOM-BLM method are very
similar to those of MSYM theory
\cite{Sabiovera:2011}\cite{Brodsky:2002ka}\cite{Brodsky:2002kn}.

In the case of the BLM method, one deals  with separate
renormalization scales for each skeleton diagram, as is done in
QED. The PMC method provides a single effective renormalization
scale which reproduces the BLM scales at NLO, even for rather
complex processes that are in our list of important projects, such
as $W$+Jets, $e^+ e^-$ annihilation, $t \bar t$ production, and
for general observables; e.g. differential cross sections,
asymmetries.

If one considers a process with high multiplicity,
then one confronts separate BLM scale for each of the multiple
skeleton diagrams; thus  the number of BLM scales will appear as
the jet multiplicity increases. The PMC method replaces these
multiple scales with an effective single scale at NLO.

We have discussed in this paper an illustration of the PMC procedure
for 3-jet production in $e^+ e^-$ annihilation where the
$n_f$ terms arise from the inclusive 4 jet cross section after IR
cancellation: these terms are included in the PMC scale with the
effect of lowering its value.

The PMC method provides the correct renormalization scale from
first principles without ambiguity or renormalization scheme
dependence.  The residual errors from the resulting conformal
series provide an accurate assessment of higher order errors. The
PMC/BLM uncertainty is zero at the order computed.The PMC is
equivalent to the standard method used to eliminate the
renormalization scale ambiguity in precision tests of QED.

The PMC method gives results which are renormalization scheme
independent at each finite order. The PMC also determines the correct
number of flavors $n_f$;  this is particularly important when
one uses a renormalization scheme which is analytic in the quark masses such as the analytic extension of the $\overline MS$ scheme~\cite{Brodsky:1998mf};  one can then
include the correct flavor threshold dependences and transitions
as one evolves the QCD coupling. The correct displacement between
the argument of the schemes is also automatically determined.

We stress that PMC  does not capture all higher-order effects. One
still has higher order corrections in the conformal series. These
can never be discovered by varying the renormalization scale,
since this variation only exposes terms proportional to the
$\beta-$function.   It is  incorrect to require the scale choice
to remove all higher order terms.   For example, in QED, the muon
anomalous moment receives a large contribution at order $\alpha^3$
from the electron-loop light-by-light insertion. This is due to
the physics of the higher-order processes --- not the running QED
coupling. It is thus incorrect to vary  the renormalization scale
to minimize the effect of higher order corrections, since the
variation of $\mu_R$ cannot expose large terms in the conformal
series.  Thus the PMC correctly and unambiguously exposes higher
order terms which are intrinsic to  physical effects, unrelated to
the QCD running coupling.

We emphasize that the PMC method  for setting the renormalization
scale gives predictions for observables which are independent of
the choice of renormalization scheme  --  a key requirement for a
valid prediction for a physical quantity.  The argument of the
running coupling in a given scheme which appears in the resulting
conformal series has the correct displacement so that the result
is scheme-independent. The number of active flavors $n_f$ in the
QCD $\beta$ function is  also correctly determined, and the
renormalization  agrees with QED scale-setting in the $N_C \to 0$
Abelian limit. Furthermore, the resulting conformal series avoids
the need for renormalon resummation.

A consistent application of the BLM/PMC procedure to B-decays,  including $B  \to X_s + \gamma$,  has been developed including resummation to all orders in the strong coupling constant.   A review and extension of this procedure is given
by Melnikov and Mitov~\cite{Melnikov:2005bx}

The PMC procedure has recently been extended to the four-loop
level,~\cite{Brodsky:2011ta} demonstrating that it provides a
consistent, systematic and scheme-independent procedure for
setting  the renormalization scales up to NNLO.\\ The explicit
application for determining the renormalization scale of
$R_{e^{+}e^-}(Q)$ up to four loops has also been
presented~\cite{Brodsky:2011ta}.


%


The PMC is the principle underlying the BLM scale-setting
procedure, a method which has been applied to many pQCD
predictions.  For example, the PMC/BLM procedure for setting the
renormalization scale is the standard method for determining the
intercept of the BFKL
pomeron~\cite{Brodsky:1998kn,Brodsky:2002ka}.

A systematic and scheme-independent procedure for setting the
PMC/BLM scales up to NNLO has also been demonstrated, including an
explicit application for determining the scale  for
$R_{e^{+}e^-}(Q)$ up to four loops~\cite{Brodsky:2011ta}. The PMC
procedure has recently been applied to the $t \bar t$
hadroproduction cross
section~\cite{Brodsky:2012sz,Brodsky:2012rj}: and the $\bar t t$
asymmetry~\cite{Brodsky:2012ik}  major  tests of the Standard
Model at colliders~\cite{Brodsky:2012sz,Brodsky:2012rj}.    The
PMC prediction for the total cross-section $\sigma_{t\bar{t}}$
agrees well with the present Tevatron and LHC data. The initial
scale-independence of the PMC prediction is found to be satisfied
to high accuracy at the NNLO level: the total cross-section
remains almost unchanged even when taking very disparate initial
scales. After PMC scale setting, the PQCD predictions  are within
$1\,\sigma$ of the CDF~\cite{Aaltonen:2011kc} and D0
measurements~\cite{Abazov:2011rq} since the relevant
renormalization scale is less than conventional estimate; the
large discrepancy of the top quark forward-backward asymmetry
between the Standard Model prediction and the data is thus greatly
reduced.

It should also be noted that the Principle of Maximum Conformality
satisfies all of the consequences of renormalization group
invariance - reflectivity, symmetry, and
transitivity~\cite{Brodsky:2012ms}. Using the PMC, all
non-conformal  in the perturbative expansion series are summed
into the running coupling, and one obtains a unique, scale-fixed,
scheme-independent prediction at any finite order. The PMC scales
and the resulting finite-order PMC predictions are both to high
accuracy independent of the choice of initial renormalization
scale, consistent with RG invariance. Moreover, after PMC
scale-setting, the residual initial scale-dependence at fixed
order due to unknown higher-order $\{\beta_i\}$-terms can be
substantially suppressed. The PMC thus eliminates a serious
systematic scale error in pQCD predictions, greatly improving the
precision of tests of the Standard Model and the sensitivity to
new physics at collider and other experiments.  Further discussion
is given in ref. ~\cite{Brodsky:2012ms}.



Clearly, the elimination of the renormalization
scheme ambiguity using the PMC will greatly increase the
precision of QCD tests and increase the sensitivity
of measurements at the LHC and Tevatron to new physics beyond the Standard Model.

\acknowledgements We thank  Xing-Gang Wu, Michael Binger, Susan
Gardner, Stefan Hoeche, Andrei Kataev, G. Peter Lepage. Al
Mueller, and Zvi Bern for helpful conversations. One of us
(L.D.G.) wishes to thank the Fondazione A. Della Riccia for
financial support and the $CP^3-$Origins Theory Group for their
hospitality.

\end{document}